\documentstyle[editedvolume,psfig]{crckapb}
\begin{opening}
  \title{Comparison between disk-like objects formed
    in hierarchical hydrodynamical simulations \protect\\
    and observations of spiral galaxies}
  \author{A. S\'aiz}
  \author{R. Dom\'inguez-Tenreiro}
  \institute{Depto.\ de F\'{\i}sica Te\'orica,
    Universidad Aut\'onoma de Madrid\\
    Madrid, E-28049, Spain}
  \author{P. B. Tissera}
  \institute{Intituto de Astronom\'{\i}a y F\'{\i}sica del Espacio\\
    Buenos Aires, 1428, Argentina}
  \author{S. Courteau}
  \institute{Univ.\ of British Columbia, Dept.\ of Physics and Astronomy\\
    Vancouver, BC, Canada V6T 1Z1}
\end{opening}
\runningtitle{Comparison between DLOs and spiral galaxies}
\begin{document}
\begin{abstract}
  We analyze
  the structural and dynamical properties of disk-like objects
  formed in fully consistent cosmological simulations which include
  inefficient star formation. 
  Comparison with data of similar observable properties
  of spiral galaxies gives satisfactory agreement, in contrast with previous
  findings using other codes.
  This suggests that the stellar formation implementation used has
  allowed the formation of disks as well as guaranteed their stability.
\end{abstract}
\section{Introduction}
We present results of a detailed comparison
between the parameters characterizing the structural and dynamical
properties of a sample of 29 simulated disk-like objects (DLOs)
and those measured
in observed spiral galaxies.
The properties we focus on are
the bulge and disk structural parameters and the rotation curves,
as they can be constrained with available data for observed spiral galaxies.
These data are taken principally from the compilation of galaxy structural
parameters of Broeils (1992), de Jong (1996) and Courteau (1996, 1997).

The DLOs have been identified in AP3M-SPH
fully consistent hierarchical hydrodynamical simulations,
realizations of a CDM flat model,
with $\Omega_{\rm b} = 0.1$, $\Lambda = 0$ and $b = 2.5$, made using
64$^3$ particles in a periodic box of comoving side 10 Mpc
(H$_0 = 50$ km s$^{-1}$ Mpc$^{-1}$),
and where an {\em inefficient\/} Schmidt law-like algorithm to model the
stellar formation processes has been implemented (see Tissera, Lambas,
\& Abadi 1997; Silk 1999).

\section{Bulge-disk decomposition}
\begin{figure}
  \centerline{%
    \psfig{bbllx=37pt,bblly=364pt,bburx=555pt,bbury=555pt,%
      file=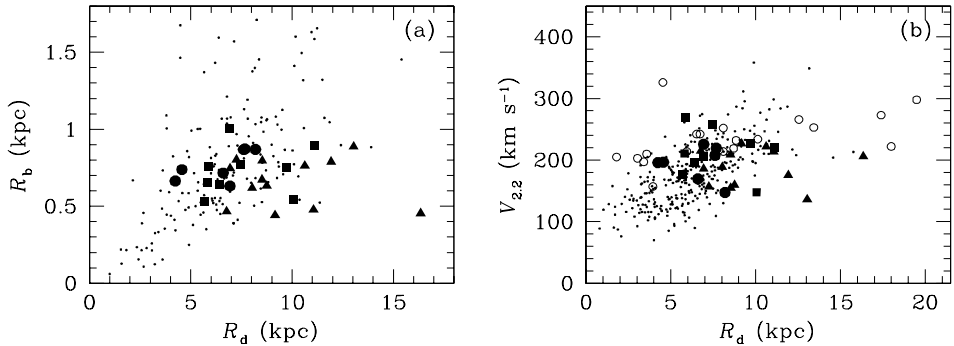,width=\textwidth,angle=0,clip=}%
    }
  \caption{The scale lengths $R_{\rm b}$ and $R_{\rm d}$ (a),
    and the $V_{2.2}$ velocities versus $R_{\rm d}$ (b).
    Filled symbols correspond to
    DLOs formed in our simulations.
    Different symbol shapes correspond to objects formed in
    different realizations.
    Dots in (a) are from 1D decompositions
    of surface brightness profiles given by Courteau, de Jong, \& Broeils
    (1996) and Courteau (1997). Dots in (b) are data from the Courteau--Faber
    (H$\alpha$ spectra) sample. Open circles are data from spirals with
    \hbox{H\,{\sc i}} rotation curves from Casertano \& van Gorkom (1991).}
  \label{bd}
\end{figure}

A surface density bulge-disk decomposition was performed
on the DLOs, using a double-exponential profile,
\begin{equation}
  \Sigma(R)=\Sigma_{\rm b}(0)\exp[-R/R_{\rm b}]+
  \Sigma_{\rm d}(0)\exp[-R/R_{\rm d}].
\end{equation}
The resulting bulge and disk scale lengths,
$R_{\rm b}$ and $R_{\rm d}$,
and their ratio $R_{\rm b}/R_{\rm d}$, are consistent with available data
(Courteau, de Jong, \& Broeils 1996; de Jong 1996;
Moriondo, Giovanelli, \& Haynes 1999). See Fig.~\ref{bd}(a).

\section{Rotation curves}
The rotation curves of the simulated objects can be constructed by
adding up in quadrature the contributions to circular rotation from
the bulge and the disk, both formed by baryonic particles,
and the dark matter halo,
\begin{equation}
  V_{\rm cir}^2(r)=V_{\rm bar}^2(r)+V_{\rm dm}^2(r).
\end{equation}

Our rotation curve analysis and shape modeling uses a
parameterization based on the spatial scale
$R_{2.2} = 2.2 \, R_{\rm d}$
and the velocity (mass) estimate 
$V_{2.2} = V_{\rm cir} (R_{2.2})$.
These parameters were compared with those measured for the
Courteau--Faber sample of bright Sb--Sc field spirals with
long-slit H$\alpha$ rotation curves,
and the Broeils
(1992) compilation of late-type spirals with extended 
\hbox{H\,{\sc i}}
rotation curves.  
Deep surface photometry is available for both samples.
In contrast
to findings
in other fully consistent
hydrodynamical simulations
(Navarro \& Steinmetz 2000 and references therein),
we find DLO $V_{2.2}$
values that are consistent with the observational data.
This is a consequence of disk formation
with conservation of the specific angular momentum $j$.
See Fig.~\ref{bd}(b).

\section{Luminous and dark matter contributions}
In real spirals, the relative contributions of baryons and dark matter
to the rotation curve cannot be inferred uniquely from the observations.
One has to postulate some hypothesis to infer the mass distribution
of the halo. The so-called `maximum disk (or maximal light) hypothesis'
(van Albada \& Sancisi 1986) assumes that the halo dark mass component
needed to fit the rotation curve should be minimum, or, equivalently,
that the contribution of the disk and bulge to the inner parts of the
rotation curves should be maximum. Considering this, one has
$(V_{\rm lum}/V_{\rm cir})_{R_{2.2}} \simeq 0.85 \pm 0.10$, at the 95\%
confidence level (Sackett 1997). The other common constraint used is that
of sub-maximal disks, obtained either by matching the vertical velocity
dispersion, scale height and scale length of a thin exponential disk
which yields the maximum disk rotation (Bottema 1993), or, independently,
showing that the Tully--Fisher relation of bright spirals is independent
of surface brightness (Courteau \& Rix 1999). Then one has
$(V_{\rm lum}/V_{\rm cir})_{R_{2.2}} \simeq 0.60 \pm 0.10$.

\begin{figure}
  \centerline{%
    \psfig{bbllx=194pt,bblly=364pt,bburx=712pt,bbury=559pt,%
      file=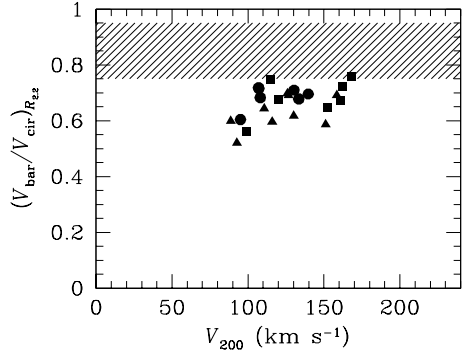,width=\textwidth,angle=0,clip=}%
    }
  \caption{The $(V_{\rm bar}/V_{\rm cir})_{R_{2.2}}$ ratios
    versus the circular velocities at virial radius, $V_{200}$,
    for the DLO sample.
    Different symbol shapes correspond to objects formed in
    different realizations.
    The shaded area shows the 95\% confidence
    interval for maximum disks.}
  \label{vbvt}
\end{figure}

The average relative contribution of baryons to $V_{2.2}$ in our sample
of DLOs can be measured directly. We find
$(V_{\rm bar}/V_{\rm cir})_{R_{2.2}}=0.67 \pm 0.07$, in very good
agreement with Bottema (1993, 1997) and Courteau \& Rix (1999).
See Fig.~\ref{vbvt}.
Most DLOs have been found to have sub-maximal disks.
This means that 
the amount of baryon mass that has ended up inside $R<R_{2.2}$
is not excessive, again as a consequence of $j$ conservation.

\section{Conclusions}
In conclusion, the comparison between the DLOs produced in our simulations and
observational data allows us to affirm that the DLOs
have counterparts
in today spiral galaxies.
This agreement suggests that the process
operating in Fall \& Efstathiou (1980) standard
model for disk formation (i.e., gas cooling and collapse with
$j$ conservation) is also at work in the
{\em quiescent\/} phases of the DLO formation in these simulations. 
However,
{\em violent\/} episodes (i.e., interactions and merger events) also occur
and play an important role in the DLO assembly (Dom\'{\i}nguez-Tenreiro,
Tissera, \& S\'aiz 1998; S\'aiz, Tissera, \& Dom\'{\i}nguez-Tenreiro 1999).
To provide the right conditions for disk regeneration
after the last violent episode of the DLO assembly, a compact central stellar
bulge is needed; this will ensure the axisymmetric character
of the gravitational potential well at scales of some kpcs at all
times, avoiding excessive $j$ losses in violent events.
A second condition is necessary: the availability of gas at low~$z$,
to form the disk.
The good match of our DLO parameters with observational
data suggests that our {\em inefficient\/} star formation
algorithm meets both requirements.
This global agreement with observations also represents
an important step towards making numerical approaches more widely used
for the study of
galaxy formation and evolution
in a cosmological framework, i.e., from primordial fluctuations.


\begin{thebibliography}{}
\bibitem[]{} Bottema R., 1993, A\&A, 275, 16;
%\bibitem[]{} Bottema R.,
  1997, A\&A, 328, 517
\bibitem[]{} Broeils A. H., 1992, Ph.D. thesis, Univ.\ Groningen
\bibitem[]{} Casertano S., van Gorkom J. H., 1991, AJ, 101, 1231
\bibitem[]{} Courteau S., 1996, ApJS, 103, 363;
%\bibitem[]{} Courteau S., 
  1997, AJ, 114, 2402
\bibitem[]{} Courteau S., Rix H.-W., 1999, ApJ, 513, 561
\bibitem[]{} Courteau S., de Jong R. S., Broeils A.H., 1996,
  ApJ, 457, L73
\bibitem[]{} de Jong R. S., 1996, A\&A, 313, 45
\bibitem[]{} Dom\'{\i}nguez-Tenreiro R., Tissera P. B., S\'aiz A., 1998,
  ApJ, 508, L123
\bibitem[]{} Fall S. M., Efstathiou G., 1980, MNRAS, 193, 189 
\bibitem[]{} Moriondo G., Giovanelli R., Haynes M. P., 1999,
  A\&A, 346, 415
\bibitem[]{} Navarro J. F., Steinmetz M., 2000, %astro-ph/0001003 preprint
  ApJ, 538, 477
%  \bibitem[]{} Rhee, M. H. 1996, Ph.D. thesis, Univ.\ Groningen
\bibitem[]{} Sackett P. D., 1997, ApJ, 483, 103
\bibitem[]{} S\'aiz A., Tissera P. B., Dom\'{\i}nguez-Tenreiro R., 1998,
  Ap\&SS, 263, 43
\bibitem[]{} Silk J., 1999, in {\sl Proceedings of the XIXth Moriond
Meeting}, eds.\ F. Hammer, T. X. Thuan, V. Cayatte, B. Guiderdoni, \&
J. Tran Thanh Van, Ed.\ Fronti\`eres, 439
\bibitem[]{} Tissera P. B., Lambas D. G., Abadi M. G., 1997, MNRAS, 286, 384
\bibitem[]{} van Albada T. S., Sancisi R., 1986, Phil.\ Trans.\ R.
  Soc.\ London Ser.\ A, 320, 447
\end{thebibliography}
\end{document}